\documentclass[10pt,preprint]{aastex}
\usepackage{graphicx}
\usepackage{natbib}
\usepackage{xspace}
\usepackage{color}

\newcommand{\irc}   		{IRC+10216\xspace}
\newcommand{\stI}     	{{\it I}\xspace}
\newcommand{\stQ}     	{{\it Q}\xspace}
\newcommand{\stU}     	{{\it U}\xspace}
\newcommand{\co}		{CO~3--2}
\newcommand{\cs}		{CS~7--6}
\newcommand{\sis}		{SiS~19--18}
\newcommand{\kms}		{~km~s$^{-1}$\xspace}

\newcommand{\cmt}		{~cm$^{-3}$\xspace}
\newcommand{\vlsr}		{~$v_{\rm LSR}$\xspace}

\begin{document}

\title{Mapping  the Linearly Polarized Spectral Line Emission around the Evolved Star IRC+10216}

\author{J.M. Girart\altaffilmark{1}}
\author{N. Patel\altaffilmark{2}}
\author{W. Vlemmings\altaffilmark{3}}
\author{Ramprasad Rao\altaffilmark{4}}

\altaffiltext{1}{Institut de Ci\`encies de l'Espai, (CSIC-IEEC), Campus UAB,
Facultat de Ci\`encies, C5p 2, 08193 Bellaterra, Catalunya, Spain;
girart@ice.cat}
\altaffiltext{2}{Harvard-Smithsonian Center for Astrophysics, 60 Garden Street,
Cambridge, MA 02138, USA}
\altaffiltext{3}{Department of Earth and Space Sciences, Chalmers University of Technology, Onsala Space Observatory, SE-439 92 Onsala, Sweden}
\altaffiltext{4}{Submillimeter Array, Academia Sinica Institute of Astronomy and Astrophysics, 645 N. Aohoku Place, Hilo, HI 96720, USA}

\begin{abstract}
We present  spectro-polarimetric observations of several molecular lines
obtained with the Submillimeter Array (SMA)\footnote{The SMA is a joint project
between the Smithsonian Astrophysical Observatory and the Academia Sinica
Institute of Astronomy and Astrophysics, and is funded by the Smithsonian
Institution and the Academia Sinica.}  toward the carbon rich AGB star
IRC+10216.  We have detected and mapped the linear polarization of the \co,
\sis\ and \cs\ lines. The polarization arises at a distance of  $\simeq 450$~AU
from the star and is blueshifted with respect the Stokes \stI. The \sis\ polarization
pattern appears to be consistent with a locally radial magnetic field configuration.
However, the \co\ and \cs\ line polarization suggests an overall complex 
magnetic field morphology within the envelope. 
This work demonstrates the feasibility of using  spectro-polarimetric
observations to carry out tomographic imaging of the magnetic field in
circumstellar envelopes.
\end{abstract}


\section{Introduction}\label{intro}

\medskip

High mass loss during the AGB phase is one of the main contributors to the
return of nucleo-synthesized material into the interstellar medium. A proper
understanding of AGB mass loss is thus crucial for the study of the chemical
evolution of the Galaxy. However, the exact mechanism or mechanisms responsible
for the AGB mass loss is still not clear. After the AGB phase, the star evolves
towards the Planetary Nebula (PN) phase. During this transition, fast winds are
launched that interact with the earlier circumstellar envelope (CSE) that was
created during the AGB. A large fraction of PNe are observed to be aspherical,
and the origin of the asphericity is attributed to the influence of a binary
companion, a disk, a magnetic field, or a combination of these. The exact onset
of asphericity is still unknown, and high angular resolution molecular line
observations indicate that several CSEs of AGB stars already display various
degrees of asymmetry. The mechanism responsible for the creation of asymmetries
is likely closely linked with that driving the mass loss and can be directly
probed by high angular resolution observations of CSEs. In particular molecular
line polarization observations are a unique tool to study potential asymmetries
in the CSE and/or determine the shape of the circumstellar magnetic field.

The critical role of magnetic fields in star formation has been probed with
direct observations of polarized dust continuum emission toward both low and
high-mass star-forming regions \citep{Girart06,GirartEtAl2009,RaoEtAl2009}.
Similar techniques are difficult to apply for AGB stars due to the need for
extremely high angular resolution ($<0.''5$) and high sensitivity.  A large
number of studies have already been made of the magnetic field induced
polarization of maser lines. These studies have revealed magnetic fields are
present throughout the entire envelope. The SiO, H$_{2}$O and OH maser
observations indicate that magnetic fields appear well ordered and the Zeeman
splitting indicates the field strengths range from several Gauss close to the
stellar surface to several mG at a few thousand AU 
\cite[e.g.,][]{Etoka04,Vlemmings05,Vlemmings11,Herpin06,Kemball09,Amiri11}.  
However, masers probe only a limited number of lines of sight through the CSE
and in most cases it is thus impossible to fully reconstruct the magnetic field
morphology throughout the envelope. Furthermore, the most abundant masers are
predominantly found around oxygen-right (M-type) limiting the available source
sample. Non-masing molecular lines are however also predicted to be linearly
polarized \citep[e.g.][]{Goldreich81, Goldreich82, Morris85} to the level of a
few percent, and can provide more extensive probes of the entire envelope.

Polarized emission has however conclusively been detected from for example 
CO in a number of star forming regions 
\citep[e.g.,][]{Girart99,LaiEtAl2003,CortesEtAl2005,ForbrichEtAl2008}. 
The molecular line polarization is due to the anisotropic
radiation field  from the central star imparting angular momentum on the
molecules or from anisotropic level populations in the molecular magnetic
substates when coupled with a magnetic field. Without a (non-radial) magnetic
field, and assuming a spherically symmetric stellar wind, linear polarization
should be radial or tangential and should only be detectable at lines of sight
away from the central star \citep{Morris85}.  Thus, when these criteria are not
met, molecular line polarization provides a unique diagnostic of magnetic field
morphology and potential non-radial asymmetries in the stellar radiation field.
Very recently, \citet{Vlemmings12} have reported the detection of CO polarized
emission toward IK Tau.

IRC+10216 (CW Leo) is a  well studied AGB star with a high mass-loss rate, 
$3\times10^{-5}$~M$_{\odot}$ yr$^{-1}$, and a terminal velocity of 15~\kms 
\citep[e.g.][]{YoungEtAl1993}.  Due to the relatively close distance of 150 pc
\citep{CrosasMenten1997},  this star provides an ideal laboratory for studies
of circumstellar chemistry \citep{Tenenbaum10, Patel11}. Unlike typical Mira
variables, there are no masers associated with this source, except perhaps a
transition of SiS which may be a weak maser \citep{Fonfria06}. The dusty
envelope of IRC+10216 shows arc-like structures in scattered light 
\citep{Mauron99} which extend over more than 1$'$ in angular radius. Closer
to the star, dust emission shows asymmetrical structures over scales of 2$''$
\citep{Menshcikov01, Menut07}.  Optical and NIR (JHK)  interferometry reveals
complex and time-varying structures on subarcsecond scales close to the star
\citep{Tuthill00, Tuthill05, Weigelt02}. Non-spherical structures were also
revealed in linear polarization maps produced by dust scattering at $0\farcs25$
resolution in H band \citep{Murakawa05}.

A previous polarization detection in IRC+10216 was reported by
\citep{Glenn97} for the CS 2--1 line.  In this letter, we present the
detection of the linear polarized emission for the \co\, \sis\ and
\cs\ lines in \irc. This is the first time that maps of the linear polarized emission
in \irc\ are presented.  

\section{Observations}

The SMA observations were taken on 2010 November 24 in the compact
configuration. The receiver was tuned to cover the 330.6-334.5 and 342.6-346.5~GHz
frequencies in the lower (LSB) and upper side band, (USB) respectively. The
phase center of the telescope was  RA(J2000.0)$=9^{\rm h}47^{\rm m}57\fs38$
and  DEC(J2000.0)$= 13\degr16\arcmin 43\farcs70$.  The correlator provided a
spectral resolution of about 0.8~MHz (i.e., 0.7~km~s$^{-1}$ at 345~GHz).  The
gain calibrators were QSOs J0854+201 and J1058+015 . The bandpass and
polarization calibrator was 3c454.3, which was observed in a parallactic angle
range of $\sim 120\arcdeg$. The absolute flux scale was determined from
observations of Titan. The flux uncertainty was estimated to be $\sim20$\%. 
The data were reduced using the MIRIAD software package \citep{Wright93}.
The SMA conducts polarimetric observations by cross correlating orthogonal 
circularly polarizations (CP). The CP is produced by inserting quarter wave 
plates in front of the receivers which are inherently linearly polarized. A 
detailed description of the instrumentation techniques as well as calibration 
issues is discussed in  \citet{Marrone08} and \citet{Marrone06}.  
In order to obtain a more accurate polarization
calibration, we solve for the leakage solution independently for the strongest
detected lines (CO and $^{13}$CO 3--2,  H$^{13}$CN 3--2, CS 7--6 and SiS
19--18) by selecting a frequency range of $\simeq 1.5$~GHz centered within
0.1~GHz with respect to the rest frequency of each line.   We found
polarization leakages between 1 and 2\% for the USB, while the LSB leakages
were between 2 and 4\%. These leakages were measured to an accuracy of 0.1\%. 
Self-calibration was performed independently for the USB and LSB on the
continuum emission of \irc.  All maps were done with Natural weighting in order
to maximize the sensitivity, which yielded a synthesized beam of
$2\farcs6\times1\farcs6$ with a position angle of  PA$\simeq0\arcdeg$ (see
caption of Figure~\ref{FigMapPol} for more specific values). Significant
polarization was only detected in the CO 3--2, CS 7--6 and SiS 7--6 lines, so
this paper presents and discuss the detection significance for these three
lines.

\begin{figure}[h]
\epsscale{1.0}
\plotone{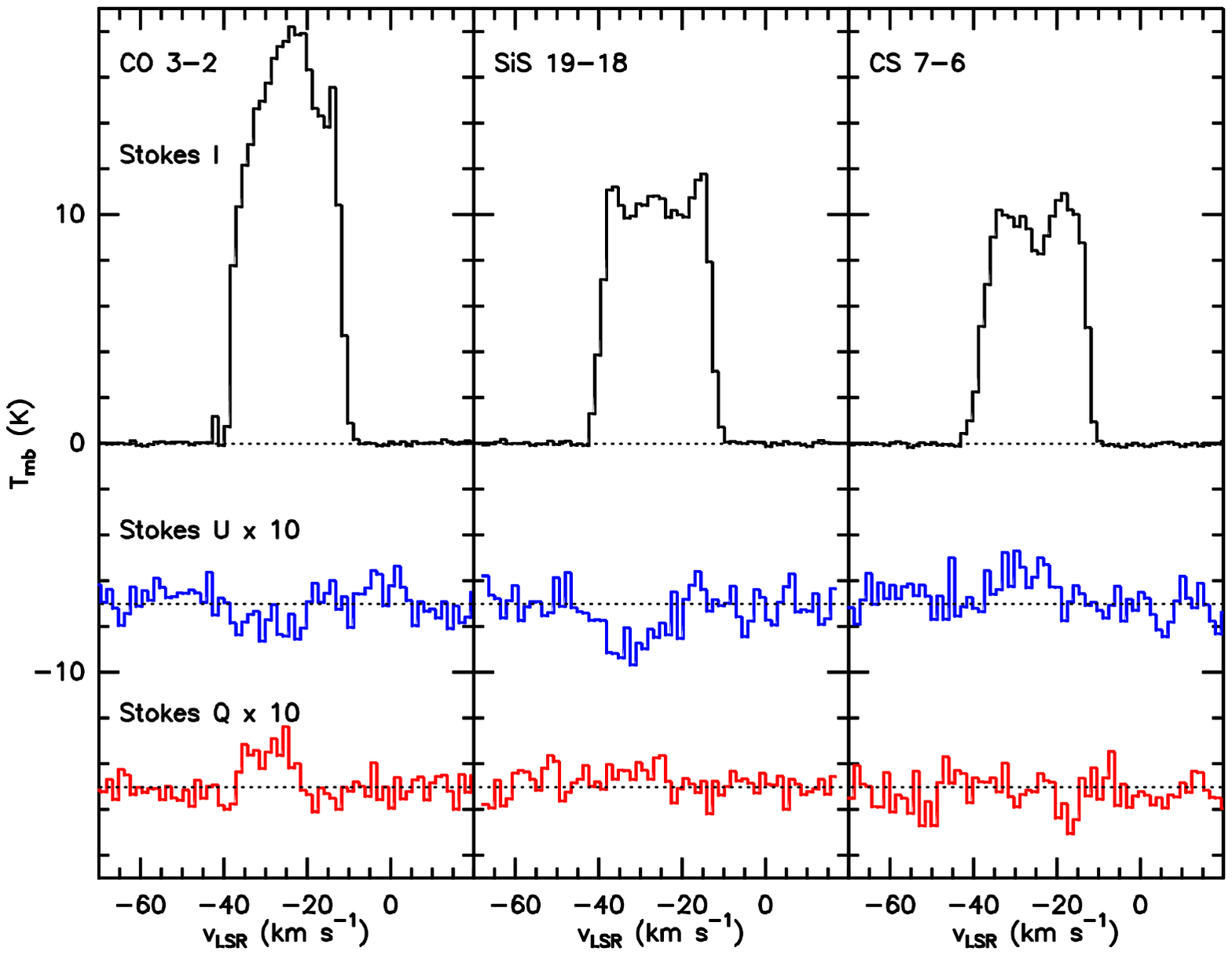}
\caption{Spectra of the Stokes $I$ (top, black line), $U$ (center, blue line)
and $Q$ (bottom, red line) emission of the CO $J$=3--2 (left panel), SiS
$J$=19--18 (central panel) and CS $J$=7--6 lines (right panel). For each line
this spectrum was taken at the position where the maximum polarized emission is
detected, after convolving the maps with a Gaussian having a FWHM of
$4''\times3''$.
}
\label{FigSpecPol}
\end{figure}

\section{Results and analysis}

\begin{figure*}
\includegraphics[angle=270,scale=0.25]{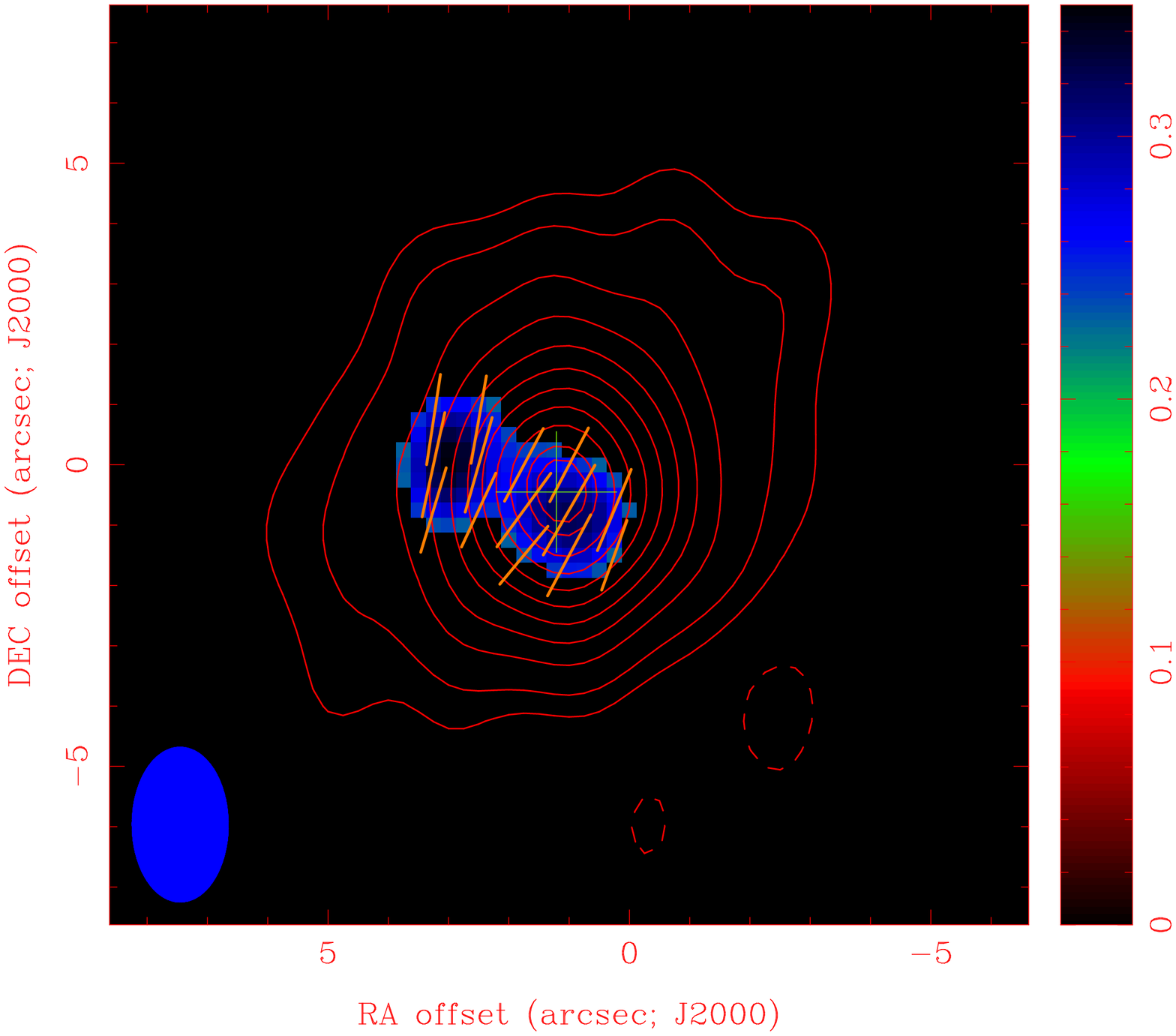}
\includegraphics[angle=270,scale=0.25]{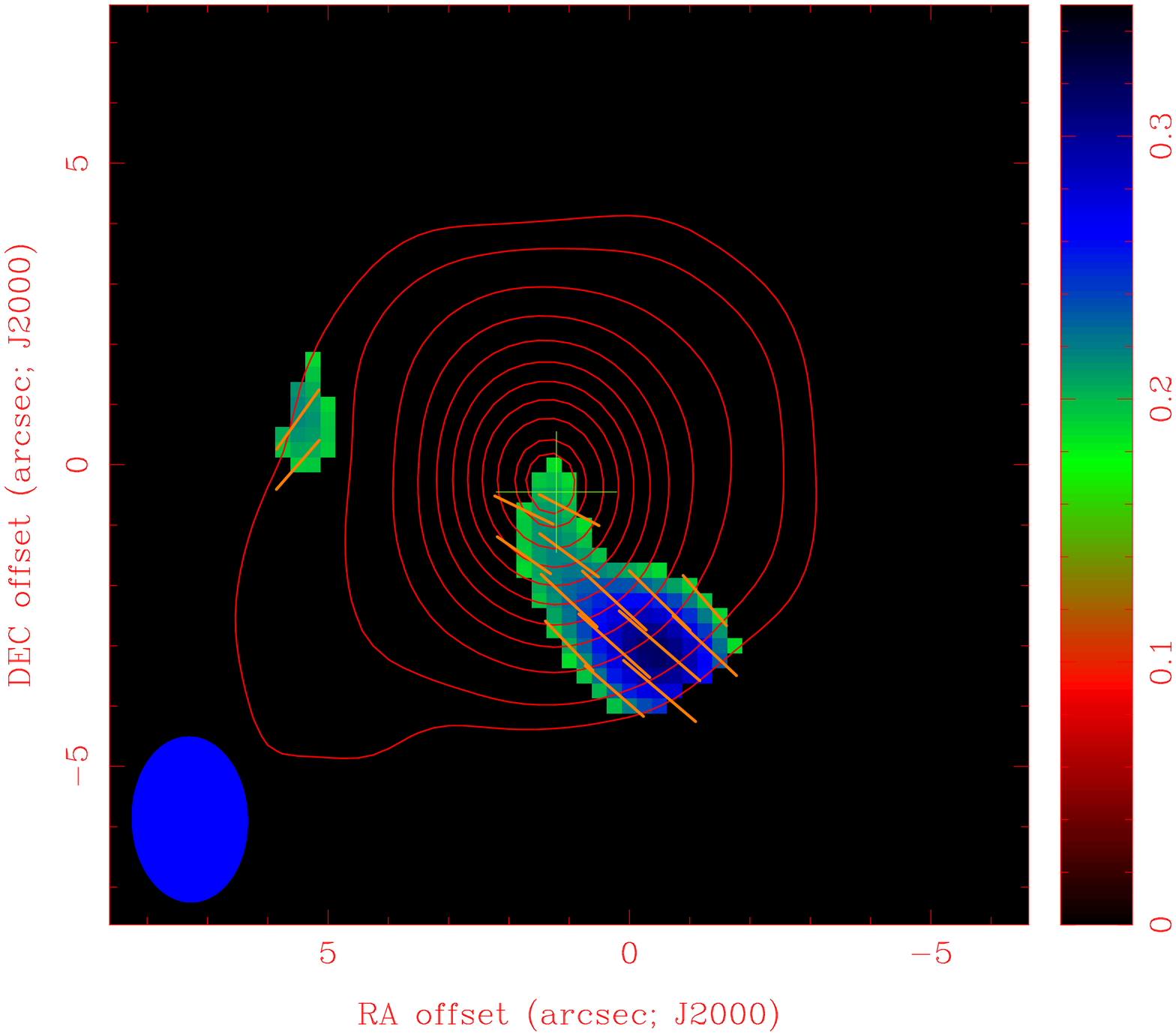}
\includegraphics[angle=270,scale=0.25]{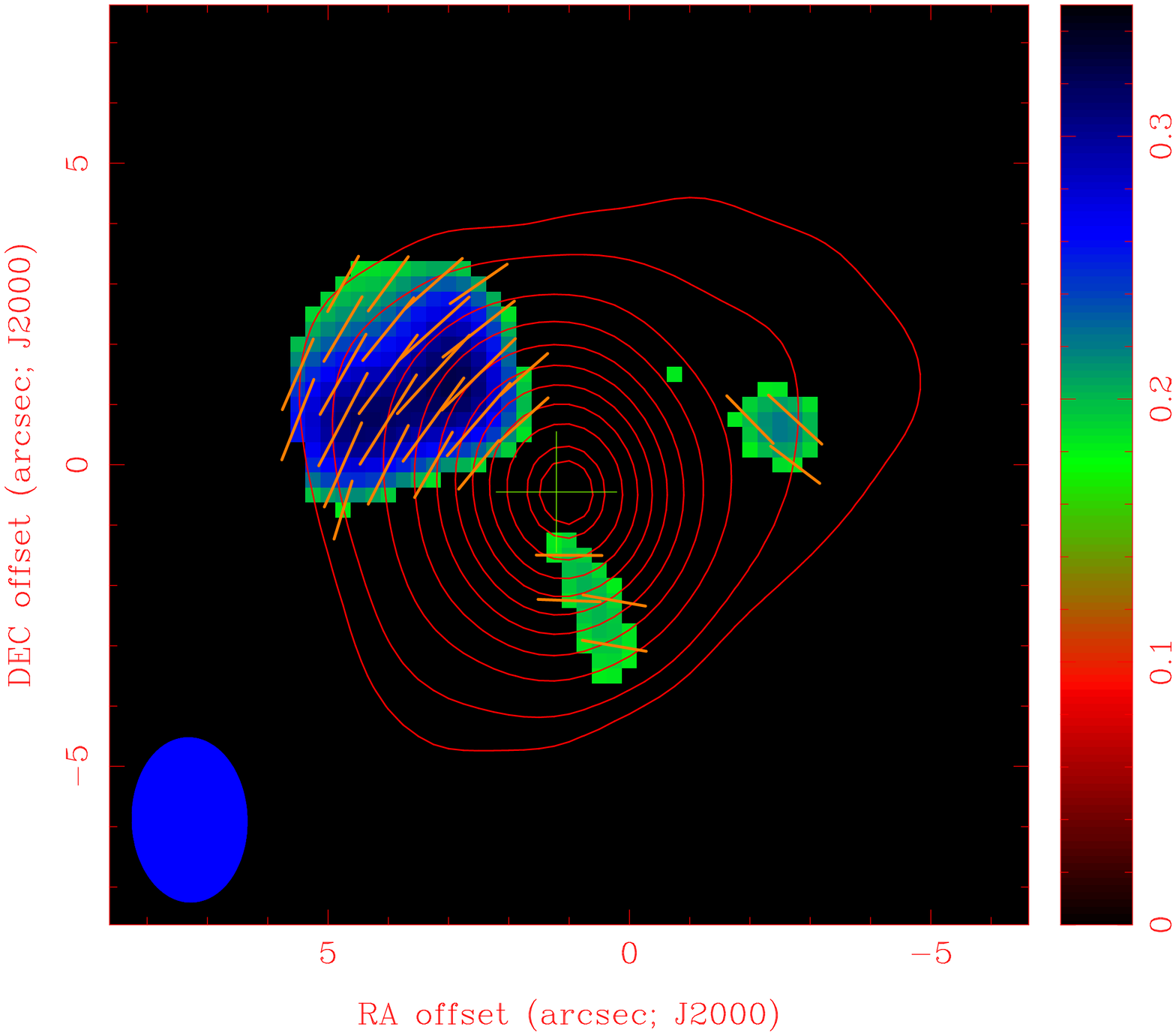}
\caption{Color image of the linearly polarized intensity of the \co\  (left
panel), \sis\  (central panel) and \cs\ lines (right panel), overlapped with
the contour maps  of the \stI\ emission for the respective lines. The orange
bars represent the polarization vectors.  The CS and CO maps show the emission
at the \vlsr\  velocity of $-29$~\kms\ averaged over 16~\kms. The SiS map shows
the emission at  \vlsr$=$-31.5~\kms\ averaged over 20~\kms.  The contour
levels  are  5, 10, 20, 30, 40, 50, 60, 70, 80, 90, 95\% of the peak intensity.
The wedge  shows the polarized intensity scale in units of Jy~beam$^{-1}$ . The
synthesized  beam is shown in the bottom left corner of each panel. 
}
\label{FigMapPol}
\end{figure*}

Figure~\ref{FigSpecPol} shows the Stokes $I$, $Q$ and $U$ obtained at the
positions of maximum polarized intensity for the \co, \cs\ and \sis\ lines.  
Figure~\ref{FigMapPol} shows the polarization maps for the emission of these
lines averaged over the velocity range that maximizes the polarized emission,
which is different for each line. 

The CO 3-2 lines is the strongest line detected in our observations (the line is 
50\% brighter than the other two lines in the shortest baselines of the
visibility plane).  This line emission is known to be spatially very extended, 
much beyond the primary
beam of the SMA  antennas \citep{Truong91}. However the SMA filters out the CO
emission that arises from structures larger than about $10''$ \citep{Girart06},
so the detected emission appears relatively compact. Indeed, the CO averaged
emission over the blueshifted component  shown in Fig.~\ref{FigMapPol} has a
relatively compact component (with a radius of $\simeq 2''$), surrounded by a
weak components that extends up $5''$ from the center at an intensity level of
about 5\% of the maximum. The CS and SiS show also a compact component with
similar dimensions, but they lack the weaker and extended component. The Stokes
\stI\  emission of the 3 lines has a similar brightness, $\simeq 120$~K for the
CS  and SiS and 150~K for CO. 

The $rms$ noise of the Stokes \stQ\ and \stU\ emission appears to slightly
increase in the channels  where the line emission is brightest. This effect is
seen in the three lines, though at different levels, from a maximum increase of
5\%, 13\% and  20\% for the CS, SiS and CO lines, respectively. This increase
is probably produced by the residual leakage   \citep[estimated to be of
$\simeq0.1$\%,][]{Marrone08} as the total intensity is very strong in the
central channels. The larger increase for the CO may be due because it is the
most extended and the brightest line (specially the shortest baselines). 
Taking into account this increase of noise, there is still significant emission
in the Stokes $Q$ and $U$ maps of the SiS at the $\simeq6$-$\sigma$  level, and
in CO and CS lines at the $\simeq5$-$\sigma$  level. The linear polarization
maps were computed by using a 3-$\sigma$ cutoff, where $\sigma$ is the rms
noise in the map where the polarization is detected. 

The CO linear polarization arises from both  the \stU\ and \stQ\ components, being
relatively bright in the later.  The CS and SiS lines show mainly polarized emission in
the \stU\ component.  Interestingly, the polarized emission in the three  lines
appears to be blueshifted with respect to the total intensity. 

In order to derive the polarization pattern in the plane of the sky, we have 
computed polarization maps with the emission averaged over the velocity  range
where the polarization intensity is detected (see Figure~\ref{FigMapPol}).  The
polarization degree at the position where it is strongest is of $\simeq2$\% for
the CO and SiS lines, and of $\simeq4$\% for the CS. The \co\ polarization
arises from two spots, one at the center of \irc\, and the other located
$\simeq 3''$ to the East. The polarization vectors are oriented  roughly
North-South, changing slightly from a position angle of  $PA = -11\arcdeg$ at
the eastern spot, to  $-25\arcdeg$ at the central spot.  The \sis\ polarized
emission arises mainly from the north-eastern  quadrant of the \irc\ envelope
(the polarization peak is located  $\simeq2\farcs6$ from the center's
envelope).  There are two other  small spots, with a polarized emission too
marginal to be further considered here. The polarization of the main component
has a mean  position angle of about $-41\arcdeg$, but the pattern of the
polarization vectors appear to form an arc, following roughly the contours of
the Stokes \stI\ emission. The  \cs\ polarization arises from the envelope's
south-western  quadrant (the polarization peak is located $\simeq2\farcs9$ from
the center's  envelope).  The polarization $PA$ pattern is quite uniform with
an averaged  value of  $\simeq 48\arcdeg$.

The SiS polarization vectors' pattern suggest a radial distribution.
Therefore,   we have compared the polarization vector direction with the
expected radial  direction (with respect to the envelope center) at the
position the vectors.   Figure~\ref{PAradial} shows the difference between the
polarization vectors  and the radial directions.  On  one hand, the SiS 
polarization vectors  are all  almost perpendicular to the radial direction,
i.e., they form a nearly perfect concentric arc-like pattern with respect to
the envelope's center.  On the other hand and despite the low polarization
statistics, this is not  the case for the CO and CS polarization vectors. 

\section{Discussion and conclusions}

The detailed analysis of population of the magnetic sub-levels in 
rotational lines show that the
highest  polarized emission is expected for volume densities similar to the 
critical density of the observed transition, and depending on the transition 
and on the molecule, the polarization can still be significant even at
densities ten  times higher \citep{Deguchi84}. This suggests that the
polarization detected in the \co\ line should arise at volume densities of  
$ \sim 10^4$~\cmt, so at the outer regions of the shell, whereas the  the
\sis\ and \cs\ polarization is expected to trace inner regions, at densities of
$\sim 10^7$~\cmt. 

One of the interesting features is that in the three lines the linear
polarization is blueshifted with respect to the total emission (this effect
is more significant in the SiS line).  Considering  that the envelope is 
expanding, this suggest that the polarized emission is being detected 
at the side of a shell facing us and with the aforementioned volume densities. 
In addition, most of the polarized
emission arises about $3''$ offset ($\simeq 450$~AU in projection) of the 
envelope's center.  Thus the optical depth is  probably 
playing an important role. Indeed, subarcsecond resolution maps in the  
IR \citep{Menut07,LeaoEtAl2006} and HCN 3--2 emission 
\citep{ShinnagaEtAl2009} show that the molecular distribution is 
asymmetrical. This suggests the anisotropy in the radiation field to be  a
cause for the polarization pattern to be not distributed spherically.
This is also in agreement with the single-dish detection of the
CS 2--1 line polarization towards the center of \irc, which suggests
that there is a non-radial polarization pattern \ \citep{Glenn97}.

\begin{figure}[h]
\epsscale{1.0}
\plotone{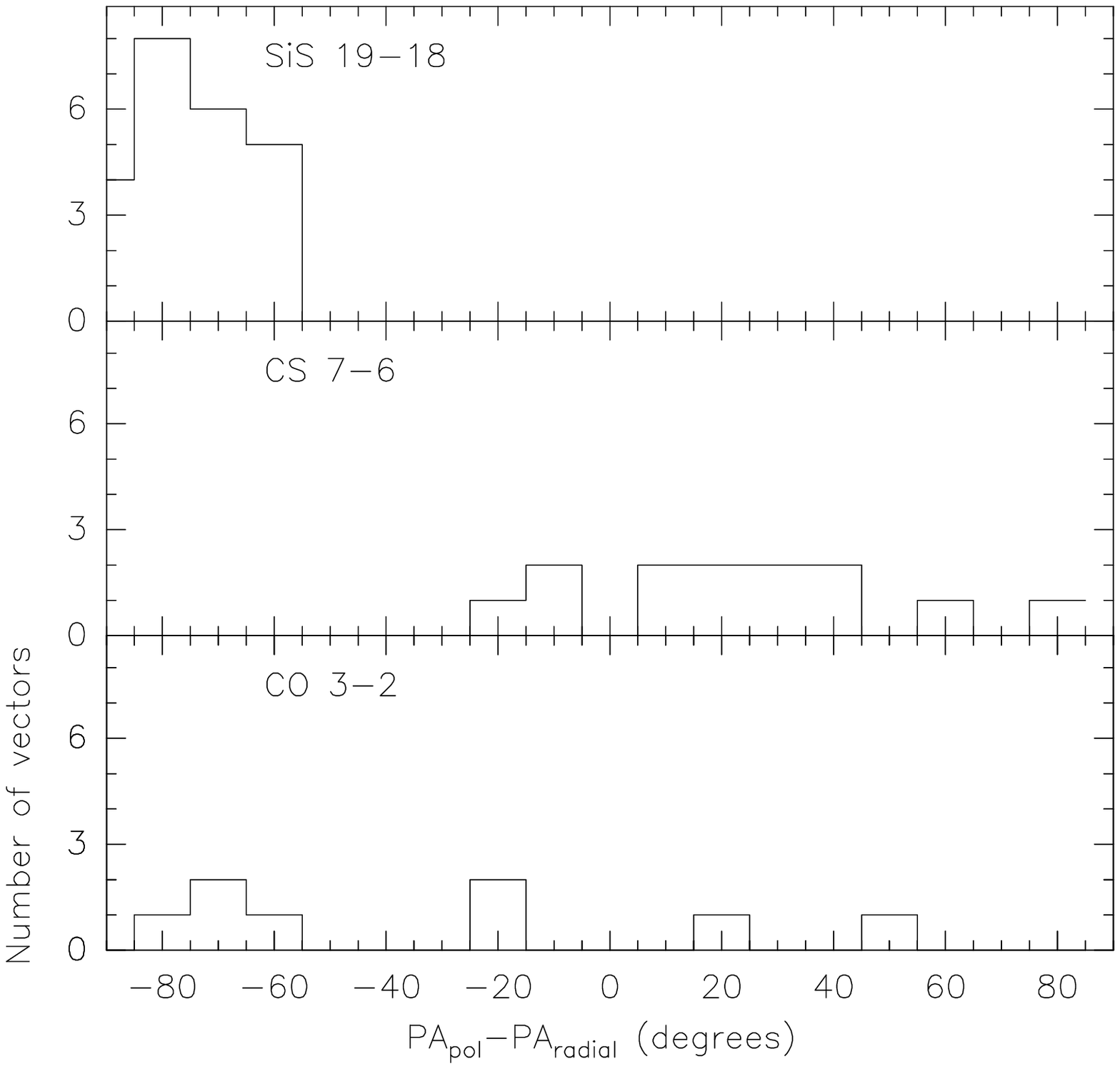}
\caption{
Distribution of the difference between the position angle of the SiS, CS and CO
polarization vectors and the radial direction with respect to the center of the
envelope. We have used a  Nyquist sample of the polarization vectors  to compute
this difference, excluding the vector closer than $\simeq 1''$ from the envelope
center.  
}
\label{PAradial}
\end{figure}

In circumstellar envelopes and for mm molecular lines the Zeeman splitting is 
much larger than the collision and spontaneous rates, even for magnetic fields 
strength of only few $\mu$G. Therefore, the polarization should be aligned 
parallel or perpendicular to the magnetic field \citep{Kylafis83, Morris85}.  
An overall radial magnetic field is expected if it is weak enough to be  energetically
irrelevant, i.e., the magnetic energy is significantly smaller than the
mechanical energy  of the stellar wind:  $B \ll (\dot{m} v_t)^{1/2} (D
\Omega)^{-1}$, where  $\dot{m}$ is the mass loss rate, $v_t$ the terminal
velocity, $D$ the distance of \irc\ and $\Omega$ the angular radius
\citep{Glenn97}. Using the measured values in \irc\ (see \S~\ref{intro}) and at
the distance where the polarization  is detected ($\simeq 3''$, 450~AU), this
condition is satisfied if $B \ll 8$~mG. CN Zeeman splitting
observations in \irc\ indicate a strength of $\simeq 9$~mG at
a larger distance  \citep[2500 AU,][]{Herpin09} . 
For a solar-type and toroidal  magnetic field
configuration  \citep[$B \propto r^{-2}$ and $r^{-1}$,
respectively;][]{Vlemmings11a}  the expected strength where the polarization is
detected would be in the 50 to 300 mG range. Therefore, the magnetic field is
strong enough to not be  radially shaped by the wind.

The measured polarized vectors can be either parallel or perpendicular to the
projected magnetic field direction in plane-of-the-sky \citep{Kylafis83}.
A proper radiative transfer analysis and a model of the physical 
conditions of the envelope (including the magnetic field configuration)
is needed to solve for this degeneracy. This applies for the measured 
CO and CS polarization maps. Nevertheless, in the region where the SiS 
polarization is detected, the polarization pattern is indicative of a radial 
magnetic field  (see Fig.~\ref{PAradial}).  Theoretical studies show that
in an expanding circumstellar envelope with a radial magnetic 
field,  the polarization will be parallel (or perpendicular) to the radial direction
if it arises at radii lower (or higher) than at certain impact parameter, $R_{J}$ 
\citep{Deguchi84, Morris85}.  $R_{J}$ is the radius where the spontaneous 
emission rate for the $J\rightarrow J-1$ transition is 
equal to the IR absorption rate.  According  to \citet{Morris85}, in
\irc\ the value  of $R_J$ for the SiS 2--1 line is $R_{2}\simeq 5 \times
10^{16}$~cm.   The spontaneous emission rate increases with J as
$J^4/(2J+1)$, and the IR absorption rate goes as $R^{-2}$, so    
$R_{19}  \simeq 2\times 10^{15}$~cm. The region
where  \sis\ polarization is detected arises at a radius of $\simeq 5 \times
10^{15}$~cm from the star.  Therefore the \sis\ polarization pattern is in
agreement with the theoretical predictions if the magnetic field is radial in
the region where SiS the polarization is detected.

In summary, we have detected and mapped the polarization pattern for the  first
time in \irc, through spectro-polarimetric observations of the \co, \sis\  and
\cs\ lines. Although, the data obtained so far lack the sensitivity to allow
us to make  specific predictions on the magnetic configuration of the \irc\
envelope, the polarization pattern measured discards that the magnetic field
configuration has a global radial pattern (this is only observed locally where the
\sis\ polarization is detected), but it possibly has a rather complex magnetic field morphology. In addition, the polarization
detection in three  different molecular lines implies that with the higher
sensitivity and angular  resolution that the Atacama Large Millimeter Array is
going to provide, it would be possible to carry out spectro-polarimetric
observations for a tomographic imaging of the magnetic field in circumstellar
envelopes. However, the ambiguity between the polarization direction with
respect to the magnetic  field direction, implies that in order to properly
relate the polarization pattern with the magnetic field, a complete radiative
transfer analysis should be made.

\acknowledgments
We thank all members of the SMA staff that made these observations possible.
JMG is supported by the Spanish MICINN AYA2008-06189-C03 and the 
Catalan AGAUR  2009SGR1172 grants. WV acknowledges support by the 
Deutsche Forschungsgemeinschaft (DFG) through the Emmy Noether 
Research grant VL 61/3-1.

\end{document}